\documentclass[useAMS]{mn2e}
\usepackage{epsfig}

\begin{document}

\def\lesssim{\mathrel{\hbox{\rlap{\hbox{\lower4pt\hbox{$\sim$}}}\hbox{$<$}}}}
\def\gtrsim{\mathrel{\hbox{\rlap{\hbox{\lower4pt\hbox{$\sim$}}}\hbox{$>$}}}}

\title[Untangling the merger history of massive black holes]
{Untangling the merger history of massive black holes with
{\it LISA}}

\author[Scott A.\ Hughes]
{Scott A.\ Hughes\thanks{E-mail: hughes@itp.ucsb.edu}\\
Institute for Theoretical Physics, University of California,
Santa Barbara CA 93106, USA}
\date{\today}
\pagerange{\pageref{firstpage}--\pageref{lastpage}}

\pubyear{2001} \volume{000}

\maketitle


\label{firstpage}

\begin{abstract}
Binary black hole coalescences emit gravitational waves that will be
measurable by the space-based detector {\it LISA} to large redshifts.
This suggests that {\it LISA} may be able to observe black holes grow
and evolve as the universe evolves, mapping the distribution of black
hole masses as a function of redshift.  An immediate difficulty with
this idea is that {\it LISA} measures certain {\it redshifted}
combinations of masses with good accuracy: if a system has some mass
parameter $m$, then {\it LISA} measures $(1+z)m$.  This mass-redshift
degeneracy makes it difficult to follow the mass evolution.  In many
cases, {\it LISA} will also measure the luminosity distance $D$ of a
coalescence accurately.  Since cosmological parameters (particularly
the mean density, the cosmological constant, and the Hubble constant)
are now known with moderate precision, we can obtain $z$ from $D$ and
break the degeneracy.  This makes it possible to untangle the mass and
redshift and to study the mass and merger history of black holes.
Mapping the black hole mass distribution could open a window onto an
early epoch of structure formation.
\end{abstract}

\begin{keywords}
gravitational waves
-- gravitation
-- black hole physics
-- cosmology: miscellaneous
\end{keywords}

\section{Introduction}
\label{sec:intro}

Evidence from quasars at redshifts $z > 5$ makes it clear that massive
black holes have existed since the universe's youth
{\cite{stern,zheng,fan}}.  In order to be powering quasars at $z \sim
5$, seed black holes with masses of $10^5\,M_\odot$ or so had to exist
at $z\sim10$ {\cite{gnedin_lisa}}.  In hierarchical formation
scenarios, these seed holes are produced by the infall and merger of
baryonic and dark matter clumps, and are then associated with dark
matter halos.  As the halos interact and merge to form larger
structures and eventually galaxies, the black holes that they contain
can merge as well [although there may be many more halo and structure
mergers than there are black hole mergers; see Milosavljevi\'c \&
Merritt (2001)].  This creates a population of coalescing massive
black holes at cosmological distances.

These coalescing black holes will be strong sources of gravitational
radiation.  It has been known for some time that such coalescences
will be detectable by the space-based gravitational-wave antenna {\it
LISA} {\cite{thorne_snow,haehnelt98,bbhI}}.  This suggests that {\it
LISA} could be used to map the distribution of black holes as the
universe evolves, tracing the growth and evolution of the massive
black hole population.  Such observations could shed light on the
merger of the black hole's host structures, providing information
about the development of structure in the universe {\cite{mhn}}.

By tracking the gravitational-wave phase as the black holes inspiral
and merge, {\it LISA} accurately measures certain mass parameters of
the binary system.  Unfortunately, the masses that {\it LISA} will
measure are redshifted.  Modulo overall amplitude, the gravitational
waves that we measure from a binary with masses $[m_1, m_2]$ at
redshift $z$ are those of a local binary with masses $[(1 + z)m_1, (1
+ z)m_2]$.  This is because any quantity $m$ with the dimension of
mass enters the binary's orbit evolution as a timescale $G m/c^3$, and
this timescale is redshifted.  Thus, the redshift and mass of a binary
black hole system are ``tangled'', greatly complicating the mapping of
the universe's black hole mass distribution.  The degeneracy between
mass and redshift must be broken or else we cannot tell the difference
between a system with mass $5\times10^4\,M_\odot$ at $z = 1$ and a
system with mass $2\times10^4\,M_\odot$ at $z = 4$.

One way to break this degeneracy would be to associate an
electromagnetic event with the coalescence.  If that event had clear
emission or absorption lines, one could directly read $z$ and thus
learn the system's mass.  In many cases, {\it LISA} will be able to
measure the luminosity distance $D$ of a source with good precision.
A measurement that gives both the luminosity distance and redshift
could be used to provide additional information about parameters such
as the Hubble constant and the cosmological constant
{\cite{cutler,markovic,wangtaylor}}.

It is likely that many more mergers will be measured with
gravitational waves than are seen electromagnetically.  Even if there
is an electromagnetic signature to the coalescence, it might be missed
because of effects such as beaming or absorption.  Fortunately,
cosmological parameters are now known well enough that we can, with
good accuracy, invert the luminosity distance as a function of
redshift, $D(z)$, to find the redshift as a function of luminosity
distance, $z(D)$.  Knowledge of the cosmological parameters allows us
to break the mass-redshift degeneracy using gravitational-wave
measurements alone, making it possible to untangle the mergers and map
the population of massive black holes.

In this paper, we examine how well the mass-redshift degeneracy can be
broken in practice.  Our eventual goal is to understand how well {\it
LISA} can measure the distribution of black hole masses as a function
of redshift.  Thus we will focus on how well {\it LISA} can measure
the mass of binary black hole systems at cosmological distances, and
on how well the luminosity distance and redshift can be determined.

It is worth noting at this point that the masses and luminosity
distance influence the gravitational waveform in very different ways.
The masses impact the binary's dynamics, particularly the orbital
frequency and inspiral rate, and thus determine the waveform's phase
evolution.  One measures the masses by tracking the phase and fitting
to a template.  If the template's phase maintains coherence with the
signal for a large number of wave cycles, the phase --- and hence the
masses --- can be measured quite accurately.  The luminosity distance,
by contrast, simply sets the overall amplitude of the wave.

Several mass parameters can be measured with gravitational waves.  The
early inspiral portion of the waveform (when the holes are
well-separated, distinct bodies) depends most strongly on the ``chirp
mass'', ${\cal M} = (m_1 m_2)^{3/5}/(m_1 + m_2)^{1/5} = \mu^{3/5}
M^{2/5}$ (where $\mu$ is the binary's reduced mass, and $M$ its total
mass).  Measuring the inspiral measures ${\cal M}$.  The inspiral also
depends (although rather more weakly) on the binary's reduced mass,
$\mu = m_1 m_2/(m_1 + m_2)$.  The reduced mass is not measured as
precisely as the chirp mass, but is often determined well enough to be
useful.  The last waves to be measured from the system's coalescence
will be the ``ringdown'' waves, emitted as the merged system relaxes
to a quiescent Kerr black hole.  These waves depend only on the final
state's mass and spin, and thus directly provide the mass $M_f$ of the
final merged remnant.

As mentioned above, the major effect of the luminosity distance is to
set the amplitude of the two gravitational-wave polarizations, $h_+$
and $h_\times$.  These amplitudes also depend upon the chirp mass, the
binary's sky position, and its orientation.  The chirp mass is
determined very precisely when the wave's phase is measured.  All that
remains to get the luminosity distance is to measure the sky position
and orientation.  These parameters are extracted by taking advantage
of {\it LISA}'s orbital motion: as the spacecraft orbit the sun, the
gravitational waveform is modulated due to the changing aspect of the
source with respect to the instrument, and due to the detector's
changing antenna pattern.  These modulations depend upon, and thus
encode, the sky position and binary orientation.  This makes it
possible to measure the luminosity distance to the source, in some
cases with a precision $\delta D/D\sim 1\%$.

The luminosity distance $D(z)$ is then inverted to give the source's
redshift, $z(D)$.  If the relevant cosmological parameters (fraction
of closure density in matter $\Omega_M$, cosmological constant
$\Lambda$, and Hubble constant $H_0$) were known with perfect
accuracy, then the redshift would be determined about as well as the
luminosity distance: $\delta z/z\simeq \delta D/D$.  In fact, each of
the cosmological parameters is presently known to about $10\%$
{\cite{wtz}}, so it turns out that the redshift can at best be
measured with about $15\%$ precision.  As future cosmological
measurements pin down the universe's geometry more precisely, the
inversion will become more precise.  By the time {\it LISA} flies (c.\
2010) it is likely that in many cases cosmological uncertainties will
not be the major source of error in determining the redshift of a
binary black hole coalescence.

To estimate how well {\it LISA} will be able to untangle binary black
hole mergers, we choose a range of system masses and randomly populate
the sky with binaries having these masses, located at redshifts $1\le
z\le 9$.  For simplicity, we focus on equal mass binaries, $m_1 =
m_2$.  We also randomly distribute the binaries' merger time within an
assumed 3 year {\it LISA} mission.  We then use a maximum likelihood
measurement formalism
{\cite{finn92,finnchernoff,cutlerflan,poissonwill}} to estimate the
precision with which each source's redshifted masses and redshift can
be measured.  For each set of redshift and system masses, we produce
the distribution of errors for the relevant parameters --- the
redshift $z$, and the redshifted chirp mass $(1 + z){\cal M}$, reduced
mass $(1 + z)\mu$, and final mass $(1 + z)M_f$.  The mass-redshift
degeneracy is thus broken for these particular mass combinations.

The remainder of this paper is organized as follows: in Sec.\
{\ref{sec:formalism}}, we briefly describe how {\it LISA} measures
gravitational waves [following the discussion of Cutler (1998)], and
summarize the formalism used to estimate parameter measurement
accuracies.  Section {\ref{sec:assumptions}} describes our models for
the gravitational waveform, the universe's geometry, and the detector
noise.  Section {\ref{sec:results}} presents our main results,
describing the accuracy with which the redshift and mass parameters
can be measured by {\it LISA}.  Using present uncertainties for our
knowledge of the cosmological parameters, we find that in many cases
{\it LISA} can measure the redshift of a system with $15-30\%$
accuracy.  The mass parameters --- particularly the chirp mass --- are
often measured much more accurately than this, suggesting that the
mass-redshift degeneracy will be broken with $15-30\%$ error.  It
should be strongly emphasized at this point that the $15\%$ lower
limit on this error is dominated by present uncertainties in our
knowledge of cosmological parameters.  If instead luminosity distance
errors dominate, it could be reduced to $5\%$ or less.

At each redshift, there is a range of system masses for which the
binary's parameters are best determined.  That is the range of masses
in which the signal lies in {\it LISA}'s most sensitive frequency
band.  As we look out to larger redshifts, this ``best mass'' becomes
smaller, since the cosmological redshift moves the relatively high
frequency signal of the smaller binary into {\it LISA}'s band.  This
is probably a very useful trend, since younger (high $z$) black holes
will tend to be smaller than older (low $z$) holes that have had more
time to grow.  Section {\ref{sec:conclude}} summarizes our major
results; we also discuss work that could improve this analysis.
Throughout, we set the speed of light $c$ and Newton's gravitational
constant $G$ to unity.  A useful conversion factor in these units is
$1\,M_\odot = 4.92\times10^{-6}\,{\rm seconds}$.  Any mass parameter
$m$ written with a $z$ subscript, $m_z$, denotes $(1 + z)m$.

\section{Gravitational-wave measurement and parameter extraction}
\label{sec:formalism}

The {\it LISA} gravitational-wave antenna {\cite{lisa}} consists of
three spacecraft arranged in an equilateral triangle orbiting the sun.
The arms of the triangle are approximately $L = 5\times10^6$ km in
length, and the triangle is inclined at an angle of $60^\circ$ to the
ecliptic.  The entire triangular configuration spins as the antenna
orbits the sun, rotating once during a single orbit (one year).  A
gravitational wave interacting with the configuration causes the
length of the three arms to oscillate.

The gravitational-wave signal is reconstructed from the time-varying
armlength data, ($\delta L_1,\delta L_2,\delta L_3$).  As described by
Cutler (1998), {\it LISA} can be regarded as two gravitational-wave
detectors.  The armlength data can be combined to produce output
equivalent to two detectors with $90^\circ$ arms; these equivalent
detectors are rotated $45^\circ$ with respect to one another.  This
``equivalent detector'' viewpoint works well when the radiation's
wavelength is greater than the armlength of the detector, but is not
so accurate when $\lambda_{\rm GW}\lesssim L$ --- high frequency
structure in the {\it LISA} sensitivity {\cite{larsonetal}} is
ignored.  We will use the equivalent detector picture throughout this
analysis, accordingly introducing some error\footnote{We expect this
error to be unimportant, since the luminosity distance and redshift
are determined from the inspiral signal, which largely accumulates at
lower frequencies.} for $f_{\rm GW} \gtrsim 0.06\,{\rm Hz}$.  The
sensitive band of the detector is taken to run from $10^{-4}\,{\rm
Hz}$ up to about $1\,{\rm Hz}$.

Following Cutler's notation, we will label the equivalent detectors
``I'' and ``II'', so the data streams are $s_I(t)$ and $s_{II}(t)$.
Each data stream consists of noise and (possibly) an astrophysical
signal.  We assume that the noises are stationary, Gaussian random
processes with the same rms value, $\langle n_I(t)^2\rangle =\langle
n_{II}(t)^2\rangle$, and that they are uncorrelated, $\langle n_I(t)
n_{II}(t)\rangle = 0$.  The two polarizations of the astrophysical
gravitational-wave, $h_+(t)$ and $h_\times(t)$, enter the datastreams
weighted by detector response functions, $F_{I,II}^+$ and
$F_{I,II}^\times$:
\begin{equation}
s_{I,II}(t) = {\sqrt{3}\over2}\left[F_{I,II}^+(t) h_+(t) +
F_{I,II}^\times(t) h_\times(t)\right] + n_{I,II}(t)\;.
\label{eq:detector_output}
\end{equation}
(The factor $\sqrt{3}/2$ enters when the outputs of the real
interferometers with arms at $60^\circ$ are converted to the
datastream of the effective interferometers, with $90^\circ$ arms.)
The response functions $F_{I,II}^{+,\times}(t)$ depend on the source's
orientation and position on the sky relative to the detector.  Because
any astrophysical signal will be at fixed position in the barycenter
frame of the solar system, the detector's motion induces modulations
in the signal's phase and amplitude.  The response functions are
written as functions of time to illustrate this orbital-motion-induced
modulation.  See Cutler (1998) for further discussion and details.

The gravitational waveform from a particular black hole binary depends
upon parameters {\mbox{\boldmath$\theta$}} which describe the system
and its evolution.  (The symbol {\mbox{\boldmath$\theta$}} stands for
a vector whose components $\theta^a$ are the distance to the system,
the masses of the black holes, their spins, the binary's position on
the sky, etc.)  Following Finn (1992), the probability that a signal
with parameters {\mbox{\boldmath$\theta$}} is present in the data
streams $s_I(t)$ and $s_{II}(t)$ is given by
\begin{eqnarray}
p({\mbox{\boldmath$\theta$}} | s_I,s_{II}) &=&
p^{(0)}({\mbox{\boldmath$\theta$}})
\exp\bigl\{-\left(H_I({\mbox{\boldmath$\theta$}}) - s_I |
H_I({\mbox{\boldmath$\theta$}}) - s_I\right)/2
\nonumber\\
& &-\left(H_{II}({\mbox{\boldmath$\theta$}}) - s_{II} |
H_{II}({\mbox{\boldmath$\theta$}}) - s_{II}\right)/2\bigr\}\;.
\label{eq:meas_prob}
\end{eqnarray}
In this equation, the inner product $(a | b)$ is
\begin{equation}
(a | b) = 4\,{\rm Re}\int_0^\infty df\,{{\tilde a}^*(f){\tilde
b}(f)\over S_h(f)}\;,
\label{eq:innerprod}
\end{equation}
where $S_h(f)$ is the spectral density of noise in the detector
(discussed further in Sec.\ {\ref{sec:assumptions}}), the superscript
$^*$ denotes complex conjugation, and ${\tilde a}(f)$ is the Fourier
transform of $a(t)$:
\begin{equation}
{\tilde a}(f) = \int_{-\infty}^\infty dt\,e^{2\pi i f t} a(t)\;.
\label{eq:fourier}
\end{equation}
The functions $H_{I,II}(t;{\mbox{\boldmath$\theta$}})$ are the
gravitational waveforms measured in detectors I and II, including the
motion-induced modulation:
\begin{equation}
H_{I,II}(t;{\mbox{\boldmath$\theta$}}) =
{\sqrt{3}\over2}\left[F_{I,II}^+(t) h_+(t;{\mbox{\boldmath$\theta$}})
+ F_{I,II}^\times(t) h_\times(t;{\mbox{\boldmath$\theta$}})\right]\;.
\end{equation}
For further discussion, see Cutler (1998).  The function
$p^{(0)}({\mbox{\boldmath$\theta$}})$ is the prior probability
distribution for the parameters {\mbox{\boldmath$\theta$}}.  It
encapsulates all information about the system's parameters known
before measurement [e.g., for each black hole, the spin magnitude
$|\vec S|$ must be less than or equal to $m^2$; see Poisson \& Will
(1995)].

The source parameters are estimated by finding the parameters
\mbox{\boldmath$\hat\theta$} that maximize
$p({\mbox{\boldmath$\theta$}}|s_I,s_{II})$; these are the ``most
likely'' parameters.  Operationally, this is done using {\it matched
filtering}: a bank of model waveforms (``templates'') that cover a
range of parameters is assembled beforehand, and the datastreams
$(s_I,s_{II})$ are cross-correlated with filters constructed from the
templates.  The parameters of the binary are estimated as those of the
template with the largest cross-correlation (i.e., the template that
``matches'' the data).  See Cutler \& Flanagan (1994),
Balasubrahmanian et al.\ (1996), and Owen (1996) for details.  The
template which maximizes the probability distribution
(\ref{eq:meas_prob}) also gives the largest value for the
signal-to-noise ratio (SNR) $\rho$:
\begin{eqnarray}
\rho &=& \sqrt{\rho_I^2 + \rho_{II}^2}\;,
\nonumber\\
\rho^2_{I,II} &=& (H_{I,II}({\mbox{\boldmath$\theta$}})|
H_{I,II}({\mbox{\boldmath$\theta$}}))\;.
\label{eq:snrdef}
\end{eqnarray}

We next estimate the errors in the measured parameters by expanding
the probability distribution function
$p({\mbox{\boldmath$\theta$}}|s_I,s_{II})$ about
${\mbox{\boldmath$\theta$}} = {\mbox{\boldmath$\hat\theta$}}$.  To do
so, we expand the inner product: denoting
$\xi_{I,II}({\mbox{\boldmath$\theta$}}) =
(H_{I,II}({\mbox{\boldmath$\theta$}}) - s_{I,II} |
H_{I,II}({\mbox{\boldmath$\theta$}}) - s_{I,II})$
{\cite{poissonwill}}, we have
\begin{equation}
\xi_{I,II}({\mbox{\boldmath$\theta$}}) =
\xi_{I,II}({\mbox{\boldmath$\hat\theta$}}) +
{1\over2} \partial_a\partial_b
\xi_{I,II}({\mbox{\boldmath$\hat\theta$}})\delta\theta^a\delta\theta^b\;.
\label{eq:innerprodexpand}
\end{equation}
Here, $\partial_a$ means partial differentiation with respect to the
parameter $\theta^a$.  In the limit of large SNR {\cite{finn92}},
$\partial_a\partial_b\xi_{I,II}/2 = (\partial_a H_{I,II} | \partial_b
H_{I,II})$, so the probability distribution becomes
\begin{equation}
p({\mbox{\boldmath$\theta$}}|s_I,s_{II}) =
p^0({\mbox{\boldmath$\theta$}})
\exp\left[-{1\over2}\Gamma_{ab}\delta\theta^a\delta\theta^b\right]\;,
\label{eq:meas_prob_expand}
\end{equation}
where
\begin{equation}
\Gamma_{ab} \equiv (\partial_a H_I | \partial_b H_I) +
(\partial_a H_{II} | \partial_b H_{II})
\label{eq:Fisher}
\end{equation}
is the Fisher information matrix.  The variance-covariance matrix
$\Sigma^{ab}$ is the inverse of this:
\begin{equation}
\Sigma^{ab} = \langle\delta\theta^a\delta\theta^b\rangle
= ({\mbox{\boldmath$\Gamma$}}^{-1})^{ab}\;.
\label{eq:covariance}
\end{equation}
The angle brackets denote an average over the probability
distribution.  Thus, the diagonal components, $\Sigma^{aa}
=\langle(\delta\theta^a)^2\rangle$, are the expected squared errors in
the parameters $\theta^a$.  Off-diagonal components describe
correlations between parameters.  It is useful to introduce the
correlation coefficient
\begin{equation}
c^{ab} = {\Sigma^{ab}\over\sqrt{\Sigma^{aa}\Sigma^{bb}}}\;.
\label{eq:correlate}
\end{equation}
This coefficient lies between $-1$ and $1$.

Equation (\ref{eq:covariance}) will be the workhorse of this analysis.
It will be used, along with models for the binary black hole
coalescence waveform and the {\it LISA} noise spectrum, in order to
estimate how well {\it LISA} will be able to measure binary
parameters, particularly the luminosity distance and the system's
masses.

\section{Model and assumptions}
\label{sec:assumptions}

Following Flanagan \& Hughes (1998), we will describe the coalescence
of the black hole binary in terms of three epochs: a slow {\it
inspiral}, in which the black holes spiral towards one another driven
by adiabatic gravitational-wave emission; a far more violent and
dynamical {\it merger}, in which the individual black holes plunge
towards one another and merge into a single body; and a final {\it
ringdown}, when the merged remnant becomes well-described as a
distorted Kerr black hole.

This characterization is rather crude.  In particular, the interface
between ``inspiral'' and ``merger'' is not very clear cut when the
members of the binary are of comparable mass.  For our purposes, this
characterization is useful because parameterized waveforms exist that
describe the inspiral and ringdown waves, and can thus can be used to
study how well we will be able to determine the masses and redshifts
of binary black hole coalescences.  At present, the merger regime
cannot be included, since its characteristics have not yet been
modeled for astrophysically interesting coalescences.  Numerical
{\cite{potsdam,harald,brandt,ggb}} and analytic
{\cite{buondamour,djs,damour}} work in this field is very active, and
hopefully will provide useful insight into the nature of the merger
and the transition from inspiral to merger by the time that {\it LISA}
begins to make observations.

\subsection{Inspiral waveform}

The inspiral waveform used here is based on the post-Newtonian
expansion of general relativity [see, e.g., Blanchet, Iyer, Will, \&
Wiseman (1996) and references therein].  We will use waveforms
computed to second-post-Newtonian (2PN) order [i.e., $(v^2/c^2)^2$
beyond the leading quadrupole result].  A very useful summary of the
2PN waveform is given in Poisson \& Will (1995).  It depends on seven
parameters: the luminosity distance to the source $D(z)$; a
coalescence time $t_c$; a coalescence phase $\phi_c$; the redshifted
chirp mass ${\cal M}_z$; the redshifted reduced mass $\mu_z$; a
spin-orbit parameter $\beta$; and a spin-spin parameter $\sigma$.

The coalescence time and phase are essentially constants of
integration, specifying the time and orbital phase of a binary at the
end of inspiral.  They are not physically interesting, but nonetheless
must be fit for in a measurement, and thus influence the accuracy with
which other parameters are measured.  The chirp mass ${\cal M} = (m_1
m_2)^{3/5}/(m_1 + m_2)^{1/5}$ is the combination of masses that most
strongly influences the gravitational-wave driven inspiral.  As we
shall see, this parameter tends to be measured to very high precision.
The reduced mass $\mu = m_1 m_2/(m_1 + m_2)$ gives the next most
important contribution to the inspiral rate.  In principle, if one
measures ${\cal M}$ and $\mu$, one can solve for the individual masses
of the holes in the binary [though it may turn out that $\mu$ is not
determined well enough for this to work well in practice
{\cite{cutlerflan}}].  The spin-orbit parameter $\beta$ describes
couplings between the spins of the black holes and the orbital angular
momentum vector.  It is given by
\begin{equation}
\beta = {1\over12}\sum_{i = 1}^2\left[113(m_i/M)^2 + 75\mu/M\right]
{\hat L}\cdot{\vec S}_i/m_i^2\;,
\label{eq:betadef}
\end{equation}
where $M = m_1 + m_2$, and ${\hat L}$ is the unit vector along the
orbital angular momentum.  The spin-spin parameter $\sigma$ likewise
describes couplings between the two spins, and is given by
\begin{equation}
\sigma = {\mu\over48M(m_1^2m_2^2)}\left[721({\hat L}\cdot{\vec S}_1)
({\hat L}\cdot{\vec S}_2) - 247 ({\vec S}_1\cdot{\vec S}_2)\right]\;.
\label{eq:sigmadef}
\end{equation}

Spin-spin and spin-orbit interactions lead to complicated precessional
motions in the binary's orbit, which in turn modulate the waveform
{\cite{acst}}.  These modulations may provide additional information
about the binary's spins and thereby reduce the effect of correlations
between various parameters.  However, including the effect of these
modulations is rather difficult.  We neglect the precession-induced
modulation of the waveform in this analysis.

In the barycenter frame of reference, the two polarizations of the
gravitational waveform described by these parameters is written
(in the frequency domain)
\begin{eqnarray}
{\tilde h}_+(f) &=& {{\cal A}\over D(z)}[1 + ({\hat L}\cdot{\hat
n})^2] f^{-7/6}\exp[i\Psi(f)]\;,\nonumber\\
{\tilde h}_\times(f) &=& -{2{\cal A}\over D(z)}({\hat L}\cdot{\hat n})
f^{-7/6}\exp[i\Psi(f)]\;.
\label{eq:inspiral_form}
\end{eqnarray}
Here, $\hat n$ is the direction vector pointing from the center of the
barycenter frame (i.e., the Sun) to the system being measured.
The amplitude is
\begin{equation}
{\cal A} = {\sqrt{5\over96\pi}}\pi^{-2/3}{\cal M}_z^{5/6}\;,
\label{eq:insp_amp}
\end{equation}
and $\Psi(f)$ is a rather complicated function of $t_c$, $\phi_c$,
${\cal M}$, $\mu$, $\beta$, and $\sigma$; see Poisson \& Will (1995),
Eq.\ (3.6).

This post-Newtonian description of $\cal A$ and $\Psi(f)$ is more
properly called the {\it restricted} post-Newtonian approximation.
The ``full'' post-Newtonian approximated waveform includes
contributions from several (in principle, all) harmonics of the
binary's orbital motion {\cite{cutlerflan}}:
\begin{equation}
h(t) = {\rm Re}\left[h_1 e^{i\Phi(t)} + h_2 e^{2i\Phi(t)} + h_3
e^{3i\Phi(t)} + \ldots\right]\;.
\label{eq:pn_full}
\end{equation}
Here, $\Phi(t)$ is the time domain orbital phase.  Each amplitude
$h_i$ itself is described by a post-Newtonian expansion; the results
rapidly get quite complicated [cf.\ Will \& Wiseman (1996), Eqs.\
(6.10) and (6.11)].  Not surprisingly, the strongest harmonic is
$h_2$, associated with the quadrupole moment of the source.  In the
restricted post-Newtonian approximation, we ignore the other harmonic
contributions to the waveform.  Further, we ignore all but the leading
``Newtonian'' order contributions to that harmonic's amplitude.
High-order post-Newtonian information is used to describe the binary's
dynamics and hence to compute the phase $\Phi$.  The Fourier transform
of this restricted post-Newtonian time domain signal then gives the
frequency domain waveform (\ref{eq:inspiral_form}).

The location vector ${\hat n}$ is given by the sky position
coordinates of the binary, $(\bar\mu_S,{\bar\phi}_S)$ (where
$\bar\mu\equiv\cos\theta$).  Likewise, the binary orientation vector
${\hat L}$ can be described using coordinates
$({\bar\mu}_L,{\bar\phi}_L)$.  As {\it LISA} orbits the sun, the
coordinates as seen by {\it LISA} continuously change, modulating the
waveform's amplitude and phase.  Simultaneously, the detector is
``rolling'', changing the profile of the arms as they ``look'' at a
source, further modulating the waveform.  For details of how this
modulation works and an elegant way to build it into the waveform, see
Cutler (1998).  These modulations make it possible to determine the
sky position of the binary to within a solid angle $\delta\Omega_S =
2\pi\delta{\bar\mu}_S\delta {\bar\phi}_S [1 -
|c^{\bar\mu_S\bar\phi_S}|]$, and the orientation of the binary to
within $\delta\Omega_L = 2\pi\delta{\bar\mu}_L\delta {\bar\phi}_L [1 -
|c^{\bar\mu_L\bar\phi_L}|]$.  If the duration of a particular
measurement is too short, these angles are determined very poorly ---
the waveform is not modulated enough.  This can severely affect the
measurement of other parameters, particularly the luminosity distance.
Note that the amplitudes ($h_1, h_2, h_3, \ldots$) defined
schematically in Eq.\ (\ref{eq:pn_full}) each depends upon these
angles in a different manner.  By ignoring all but $h_2$, the
restricted post-Newtonian approximation throws away information that,
in principle, could be used to improve gravitational-wave measurement
accuracy.  This is a natural point for an improved follow up to this
analysis, as will be discussed further in Sec.\ {\ref{sec:conclude}}.

We terminate the inspiral when the binary's members are separated by a
distance $6M$; this very roughly corresponds to the point at which the
post-Newtonian expansion ceases to be accurate.  The
gravitational-wave frequency at this point is
\begin{eqnarray}
f_{\rm gw}(r = 6M) &=& {2f_{\rm orb}(r = 6M)\over(1 + z)}\nonumber\\
&=& {2\Omega_{\rm orb}(r = 6M)\over2\pi(1 + z)}\nonumber\\
&=& {1\over(1+z)\pi}\sqrt{M\over(6M)^3}\nonumber\\
&=& {6^{-3/2}\over\pi M_z}\nonumber\\
&\simeq& 0.04\,{\rm Hz}\left({10^5\,M_\odot\over M_z}\right)\;.
\label{eq:f_end_insp}
\end{eqnarray}

The total number of inspiral parameters is eleven: the four position
and orientation coordinates, the distance to the source, the constants
of integration $t_c$ and $\phi_c$, and four combinations of the
binary's masses, spins, and orbital angular momentum.  In Sec.\
{\ref{sec:results}}, we determine the accuracy with which these
parameters can be measured for a wide variety of interesting systems
using the parameter measurement formalism discussed in Sec.\
{\ref{sec:formalism}}, particularly Eq.\ (\ref{eq:covariance}).  Note
that we confine our discussion to the masses, the luminosity distance,
and the redshift, since our primary interest is to understand what
{\it LISA} can say about mapping the black hole mass distribution.  In
the course of this analysis we also fit for and estimate the errors on
all the other parameters discussed in this section.

\subsection{Ringdown waveform}

In some cases, the waves from the final ringdown will be measurable as
well.  These waves are emitted as the system settles down to the
stationary Kerr black hole solution.  They are of relatively high
frequency, so a coalescence with a very interesting inspiral may have
a ringdown that is entirely lost in high frequency noise.  Likewise,
some systems have inspirals that are overwhelmed by low frequency
noise, but emit ringdown waves right in the band of maximum
sensitivity.

Weak distortions of Kerr black holes can be decomposed into spheroidal
modes, with spherical-harmonic-like indices $l$ and $m$
{\cite{leaver}}.  Each mode oscillates with a unique frequency
$f_{lm}$ and damping time $\tau_{lm}$, generating gravitational waves
whose form is a damped sinusoid.  The frequency and damping time
depend only on the mass and spin of the black hole.  Measuring
ringdown waves thus measures the mass and spin of the coalesced
system.  This provides additional information about the black hole
mass distribution.  It can also be combined with the chirp mass $\cal
M$ and reduced mass $\mu$ measured during the inspiral to infer the
initial masses $(m_1, m_2)$ of the binary's members.  (Because a
fraction of the system's mass is radiated away during the merger and
ringdown, some systematic error is necessarily introduced when mass
determined from the ringdown is combined with masses determined from
the inspiral.)

The ringing waves emitted at the endpoint of binary coalescence will
presumably be dominated by the $l = m = 2$ mode.  This is a bar-like
mode that propagates about the equator in the same sense as the hole's
spin.  It is likely to dominate at late times because the coalescing
system has a shape that nearly mimics an $l = m = 2$ distortion (so
that it should be preferentially excited), and also because this mode
is more long-lived than any other.  A more detailed understanding of
the merger epoch is needed to better assess the likely mixture of
modes at the end of coalescence.

A good fit to the frequency $f_{22} \equiv f_{\rm ring}$ and quality
factor $Q\equiv\pi f_{\rm ring}\tau$ is {\cite{echeverria}}
\begin{eqnarray}
f_{\rm ring} &=& {1\over2\pi M_z}\left[1 - 0.63(1 -
a)^{3/10}\right]\;,
\label{eq:fring}\\
Q &=& 2(1 - a)^{-9/20}\;,
\label{eq:Qring}
\end{eqnarray}
where $a = |{\vec S}|/M^2$ is the dimensionless Kerr spin parameter.
A merged remnant with mass $10^5\,M_\odot$ at $z = 1$ would emit
ringdown waves somewhere in the band from $f = 0.06\,{\rm Hz}$ ($a =
0$) to $f = 0.16\,{\rm Hz}$ ($a = 1$).  The ringdown frequency is
always quite a bit higher than inspiral frequencies.

As mentioned above, the time domain gravitational waveform for the
ringdown waves are damped sinusoids.  They can be written
\begin{eqnarray}
h_+(t) &=& {\cal A}_+\exp(-\pi f_{\rm ring} t/Q)\cos(2\pi f_{\rm ring}
t + \varphi)\;,
\nonumber\\
h_\times(t) &=& {\cal A}_\times\exp(-\pi f_{\rm ring} t/Q)\sin(2\pi
f_{\rm ring} t + \varphi)\;.
\label{eq:ringwave}
\end{eqnarray}
It is not easy to estimate the polarization amplitudes ${\cal
A}_{+,\times}$; they will depend upon the detailed evolution of the
merger epoch, as well as variables such as the orientation of the
final merged remnant.  A reasonable hypothesis is that their ratio
follows the ratio of the inspiral polarization amplitudes:
\begin{eqnarray}
{\cal A}_+ &=& {\cal A}_{\rm ring}\left[1 + ({\hat L}\cdot{\hat
n})^2\right]\;,
\nonumber\\
{\cal A}_\times &=& -2{\cal A}_{\rm ring}({\hat L}\cdot{\hat n})\;.
\label{eq:pol_amplitudes}
\end{eqnarray}
We set the overall amplitude ${\cal A}_{\rm ring}$ by requiring that
the ringdown radiate some fraction $\epsilon$ of the system's total
mass [see Fryer, Holz, \& Hughes (2001), Sec.\ 2.5].  The result is
\begin{equation}
{\cal A}_{\rm ring} = {1\over D(z)}\sqrt{5\epsilon M_z\over4\pi f_{\rm
ring} Q}\;.
\label{eq:amplitudes}
\end{equation}
It seems likely that $\epsilon$ will vary rather strongly depending
upon the constituents of the binary, particularly the black holes'
spins and spin orientations.  For concreteness, we will use $\epsilon
= 1\%$ in all calculations.  This is in accord with the fraction of
system mass radiated in recent numerical simulations [see, for
example, Baker et al.\ (2001) and references therein].  The phase
$\varphi$ essentially tells us the configuration of the merged remnant
when the $l = m = 2$ mode dominates its dynamics.  We will take it to
be randomly distributed over mergers.

Whereas the inspiral waveform may be measured by {\it LISA} over the
course of several months or years, the ringdown for any source
considered here will last no more than a few hours (and in many cases,
only a few minutes).  This can be seen by using Eqs.\ (\ref{eq:fring})
and (\ref{eq:Qring}) with $\tau = Q/\pi f_{\rm ring}$.  The
motion-induced modulation of the ringdown waveform may therefore be
ignored.  The ringdown provides no information about the binary's
location on the sky --- the only characteristics of the binary we are
likely to learn from the ringdown are the mass and spin of the remnant
hole.

Because the ringdown waves are fairly narrow in frequency space, we
may make some useful approximations when computing the expected SNR
from a ringdown measurement.  Suppose the ringdown begins at time
$T_{\rm ring}$.  Define $H_{I,II}(t) = \sqrt{3}[F_{I,II}^+(T_{\rm
ring}) h_+(t - T_{\rm ring}) + F_{I,II}^\times(T_{\rm ring})
h_\times(t - T_{\rm ring})]/2$ (since we ignore detector motion, the
response functions are only evaluated at $t = T_{\rm ring}$).  Then,
we have
\begin{eqnarray}
\rho^2_{I,II} &=& 4\int_0^\infty df\;{|{\tilde H}_{I,II}(f)|^2\over S_h(f)}
\nonumber\\
&=& 2\int_{-\infty}^\infty df\;{|{\tilde H}_{I,II}(f)|^2\over
S_h(|f|)}
\nonumber\\
&\simeq& {2\over S_h(f_{\rm ring})}\int_{-\infty}^\infty df\;
|{\tilde H}_{I,II}(f)|^2
\nonumber\\
&\simeq& {2\over S_h(f_{\rm ring})}\int_{-\infty}^\infty dt\;
H_{I,II}(t)^2
\nonumber\\
&\simeq& {2\over S_h(f_{\rm ring})}\int_{T_{\rm ring}}^\infty dt\;
H_{I,II}(t)^2
\label{eq:ringsnr}
\end{eqnarray}
On the third line, we have used the fact that most of the signal power
is at $f = f_{\rm ring}$ to pull the noise out of the integral, on the
fourth line we have used Parseval's theorem to go from a frequency
domain to a time domain integral, and on the fifth line we have used
the fact that ringdown waves by definition are zero before $T_{\rm
ring}$.  This final integral is simple to evaluate using Eq.\
(\ref{eq:ringwave}).

We could now apply the full parameter estimation formalism discussed
in Sec.\ {\ref{sec:formalism}} to see how well the mass $M$ and spin
$a$ of the remnant black hole can be determined.  Such an analysis has
in fact already been performed {{\cite{finn92}}}.  Because the
detector motion has no important impact, Finn's results carry over
directly to this analysis.  In particular, the final mass is
determined to an accuracy
\begin{equation}
{\delta M_{f,z}\over M_{f,z}} = {{\cal F}(a)\over\rho} \simeq
{2\over\rho} = {2\over\sqrt{\rho_I^2 + \rho_{II}^2}}\;.
\label{eq:Mmeasure}
\end{equation}
The function ${\cal F}(a)$ slowly varies from roughly 2.5 for a
Schwarzschild black hole to 0.5 for an extreme Kerr hole.  Since the
goal of this analysis is a simple estimate, and also since the initial
conditions for ringing waves from coalescences are not well known, we
consider approximating ${\cal F}(a)\simeq2$ to be as accurate as is
warranted.

\subsection{Cosmological model}

The cosmological parameters enter this analysis through the need to
convert between redshift and luminosity distance.  We will assume a
flat cosmology, with matter and cosmological constant contributions to
the total density given by $\Omega_\Lambda = 0.65$ and $\Omega_M = 1
-\Omega_\Lambda = 0.35$.  These choices are in accord with recent
observational evidence (e.g., Netterfield et al.\ 2001).  The
luminosity distance to a source at redshift $z$ is given by [Hogg
(1999) and references therein]
\begin{equation}
D(z) = {(1 + z)c\over H_0}\int_0^z {dz'\over E(z')}\;,
\label{eq:lumdist}
\end{equation}
where
\begin{equation}
E(z) = \sqrt{\Omega_M (1 + z)^3 + \Omega_\lambda}\;.
\label{eq:Eofz}
\end{equation}
The Hubble constant $H_0 = 100\,h_0\,\mbox{km/(Mpc s)}$; we assume
$h_0 = 0.65$ {\cite{wtz}}.

This expression is very easy to invert numerically; we do so with a
simple bisection, obtaining $z(D)$.  For a particular
gravitational-wave measurement, we get $D$ with some error $\delta D$.
This error, and also errors in the cosmological parameters
$\Omega_\Lambda$ and $h_0$, mean that $z$ is measured to a precision
$\delta z$:
\begin{equation}
\delta z = {\partial z\over\partial D}\left[\delta D^2 +
\left(\partial D\over\partial\Omega_\Lambda\right)^2
\delta\Omega_\Lambda^2 +
\left(\partial D\over\partial h_0\right)^2\delta h_0^2\right]^{1/2},
\label{eq:redshift_error}
\end{equation}
where
\begin{eqnarray}
{\partial z\over\partial D} &=&
\left({\partial D\over\partial z}\right)^{-1}\,,
\nonumber\\
{\partial D\over\partial z} &=& {c\over H_0}\left[{(1 + z)\over E(z)} +
\int_0^z {dz'\over E(z')}\right]\;,
\nonumber\\
{\partial D\over\partial\Omega_\lambda} &=& {(1 + z)c\over2H_0}
\int_0^z dz'\,{\left[(1 + z')^3 - 1\right]\over E(z')^3}\;,
\nonumber\\
{\partial D\over\partial h_0} &=& -{D(z)\over h_0}\;.
\label{eq:lumdist_derivs}
\end{eqnarray}
Note that $\delta\Omega_M$ is not included since we have assumed that
the universe is precisely flat, and therefore $\delta\Omega_M =
-\delta\Omega_\Lambda$.  Following the discussion in Wang, Tegmark, \&
Zaldarriaga (2001), we will put $\delta\Omega_\Lambda = 0.1$ and
$\delta h_0 = 0.1$.  When $\delta D/D \ll 10\%$, $\delta z$ is
dominated by $\delta\Omega_\Lambda$ and $\delta h_0$; in these cases,
it typically turns out that $\delta z/z \simeq 15\%$.  We will also
look at redshift measurement accuracy assuming that
$\delta\Omega_\Lambda = 0 = \delta h_0$; this demonstrates how well
the redshift could be measured in principle if the universe's geometry
were known perfectly.

This assumed cosmology is adequate for the purposes of this paper.  As
our knowledge of the universe's geometry improves, this prescription
can be readily generalized (for example, if the ``dark energy'' turns
out to have an equation of state other than that of a cosmological
constant).

\subsection{Detector noise}

Finally, we need a model for the noise in the {\it LISA} detector to
proceed.  As discussed in Sec.\ {\ref{sec:formalism}}, we assume that
the noise in each of the effective detectors is a stationary, Gaussian
random process.  We also assume that their noises are uncorrelated.
As discussed by Cutler (1998), one can combine the data streams in
such a way that this second assumption is true by construction.

The detector noise is characterized by the spectral density $S_h(f)$.
This quantity is the Fourier transform of the autocorrelation of the
time domain noise:
\begin{eqnarray}
S_h(f) &=& 2\int_{-\infty}^\infty d\tau\, C_n(\tau)e^{2\pi i f\tau}\;,
\nonumber\\
C_n(\tau) &=& \langle n(t) n(t + \tau) \rangle\;,
\label{eq:specdensdef}
\end{eqnarray}
where angle brackets denote ensemble averaging.  We assume that the
spectral density is the same for each of the effective detectors.

{\it LISA}'s datastream will include instrumental noise intrinsic to
the detector, and confusion noise arising primarily from the large
number of gravitational wave producing white dwarf binaries in the
galaxy.  Especially at low frequencies, one cannot resolve these
binaries individually --- for a mission lasting $T\sim{\rm
several}\,10^7\,{\rm seconds}$, there could be as many as $10^3$
binaries contributing power in a single frequency bin $\delta f \sim
1/T$ {\cite{cutler}}.  The unresolved white dwarf binaries constitute
a stochastic background that, from the perspective of measuring black
hole binary waves, is a source of noise.

A good fit {\cite{phinneyprivate}} to the projected instrumental
noise for {\it LISA} {\cite{folkner}} is given by
\begin{eqnarray}
S^{\rm inst}_h(f) &=& \Bigl\{s_1[1 + f/(0.01\,{\rm Hz})] + s_2/f^2
\nonumber\\
&+& s_3/f^{5/2} + s_4(10^{-4}\,{\rm Hz}/f)^{20}\Bigr\}^2\;,
\label{eq:instrumental}
\end{eqnarray}
where
\begin{eqnarray}
s_1 &=& 4\times10^{-21}\,{\rm Hz}^{-1/2}\;,\nonumber\\
s_2 &=& 3\times10^{-26}\,{\rm Hz}^{3/2}\;,\nonumber\\
s_3 &=& 5\times10^{-28}\,{\rm Hz}^2\;,\nonumber\\
s_4 &=& 1\times10^{-16}\,{\rm Hz}^{-1/2}\;.
\end{eqnarray}
Notice that this noise shoots up very rapidly at frequencies less than
$10^{-4}\,{\rm Hz}$.  In fact, {\it LISA} could have good sensitivity
at lower frequencies than this [cf.\ Fig.\ 5 of Larson, Hiscock, \&
Hellings (2000)].  The very low-frequency sensitivity will be limited
by {\it LISA}'s ability to maintain ``drag-free'' motion on long
timescales (that is, motion driven solely by gravity, without
interference from external forces due to, for example, the solar
wind).  In cutting off {\it LISA}'s sensitivity at $10^{-4}\,{\rm
Hz}$, we assume that drag-free behavior will be difficult to maintain
on timescales longer than about 3 hours.

The galactic confusion noise is well described by
\begin{eqnarray}
S^{\rm gal}_h(f) &=& 5\times10^{-44}\,{\rm Hz}^{-1}\left({1\,{\rm
Hz}\over f}\right)^{7/3}\nonumber\\
&\times&\left[1 - \exp(-\delta f dN/df)\right]\;.
\label{eq:wdgal}
\end{eqnarray}
The factor in square brackets gives the fraction of frequency bins
near $f$ that have a contribution from galactic white dwarf binaries.
This fraction decreases with increasing frequency --- the population
of white dwarf binaries thins out in frequency space at high $f$.
This is because the inspiral rate $df/dt$ grows with $f$, so a
population that is initially clustered near some frequency spreads out
as it evolves.  In this equation, we put $\delta f = 3/T$, where $T$
is the duration of the {\it LISA} mission, and the factor 3 roughly
accounts for the smearing of a white dwarf binary's signal due to
detector motion.  The function
\begin{equation}
{dN\over df} = 2\times10^{-3}\,{\rm Hz}^{-1}\left(1\,{\rm Hz}\over
f\right)^{11/3}
\label{eq:dNdf}
\end{equation}
is the number of white dwarf binaries per unit gravitational-wave
frequency.  The fit (\ref{eq:wdgal}) was provided by Phinney
{\cite{phinneyprivate}}; the prefactor $5\times10^{-44}$ and the value
of $dN/df$ correspond closely to the results presented in Webbink \&
Han (1998).

At high enough frequencies, there will be on average fewer than one
white dwarf binary per bin.  The confusion noise goes from a smooth
continuum to a series of non-confused lines.  For galactic binaries,
these lines should be strong enough that they can be fit and
subtracted from the data stream, reducing the overall noise.  A simple
estimate of the resulting noise {\cite{phinneyprivate}} is
\begin{eqnarray}
S_h^{\rm inst + gal}(f) = {\rm min}
\biggl[S_h^{\rm inst}(f)/\exp(-\delta f dN/df),
\nonumber\\
S_h^{\rm inst}(f) + 5\times10^{-44}\,{\rm Hz^{-1}}\left(1\,{\rm
Hz}/f\right)^{7/3}\biggr]\;.
\label{eq:sum}
\end{eqnarray}
The factor $\exp(-\delta f dN/df)$ is the fraction of empty bins.  For
the total {\it LISA} noise, we take (\ref{eq:sum}) plus a contribution
from extragalactic binaries {\cite{folkner}}:
\begin{equation}
S_h^{\rm ex.\ gal}(f) = 1.1\times10^{-46}\,{\rm Hz}^{-1}\left(1\,{\rm
Hz}\over f\right)^{7/3}\;.
\label{eq:wdexgal}
\end{equation}

Finally, it should be noted that Phinney has recently re-examined the
issue of confusion limited backgrounds in very general terms
{\cite{phinney_mnras1}}, and will soon produce a paper pointing out
some overlooked sources of confusion noise {\cite{phinney_mnras2}}.
These noises are not included in this analysis, but could be very
easily.  Their main effect will be to augment the low-frequency noise,
thus reducing the signal-to-noise ratio.  The accuracy with which
parameters can be measured will be reduced as well, but very likely
that reduction will nearly scale with the reduction in SNR.

\section{Results}
\label{sec:results}

To determine how well {\it LISA} will be able to measure masses and
redshift, we have performed a large number of Monte Carlo simulations
of binary black hole coalescence measurement.  We choose binary masses
and a redshift, and then randomly distribute 100 such binaries over
sky position and orientation.  We also randomly distribute the binary
members' spins, and the magnitude of the final merged hole's spin.
(Of course, the final spin should be found by conserving the initial
spin and orbital angular momentum of the system, less that which is
lost to radiation.  The details of this are complicated and depend to
some degree on poorly understood physics.  For the purpose of
estimating how well the final system mass is determined, randomly
choosing the final spin should be fine.)  Finally, we assume that the
{\it LISA} mission lasts three years, and we uniformly distribute the
coalescence time of the binary during that time.  As a consequence,
the total observation time may vary quite a bit among inspirals in a
particular run, which can have a large impact on parameter
determination.  The routine {\tt ran2()} {\cite{numrec}} was used to
generate random numbers.

\begin{figure}
\epsfig{file = 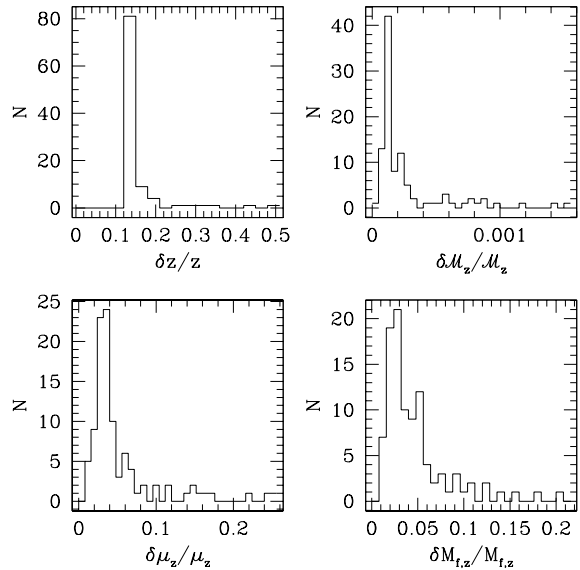, width = 8cm}
\caption{Distribution of errors in $z$, ${\cal M}_z$, $\mu_z$, and
$M_{f,z}$ for {\it LISA} measurement of a binary with $m_1 = m_2 =
10^5\,M_\odot$ at $z = 1$.  The typical inspiral SNR is about 1000;
the ringdown SNR is around $65$.  This mass is nearly optimal at this
redshift for determining the binary's parameters.  The error in $z$ is
dominated by errors in the cosmological parameters; the luminosity
distance is actually determined quite precisely for this source (cf.\
Fig. {\ref{fig:z1m1e5D}}).}
\label{fig:z1m1e5}
\end{figure}

\begin{figure}
\epsfig{file = 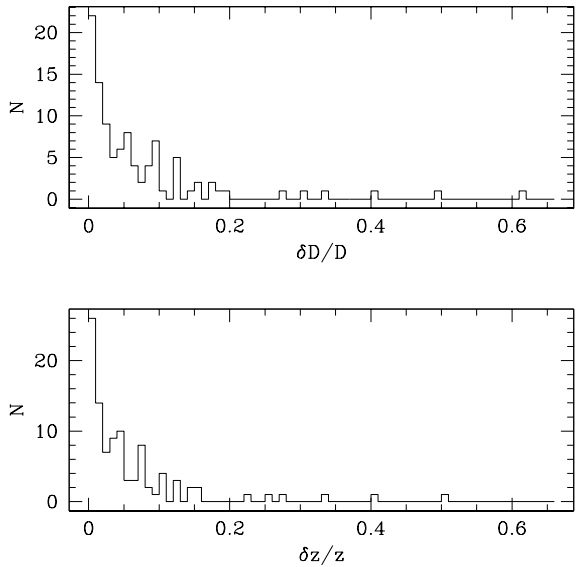, width = 8cm}
\caption{The top panel shows the distribution of error in luminosity
distance $D$ for {\it LISA} measurement of binaries with $m_1 = m_2 =
10^5\,M_\odot$ at $z = 1$; the bottom panel shows how well the
redshift could be measured if cosmological parameters were known
perfectly.  In both cases, the peak of the distribution is at a
relative error near $1\%$.  This indicates that in many cases the
mass-redshift degeneracy will be broken with very good precision when
cosmological parameters are better determined.  Note, though, the very
large tail in the distribution, extending out to relative error of
about $60\%$.  The distance to the source is poorly determined when a
source's sky position is poorly determined.  This typically happens if
the observation time is short --- the merger occurs near the beginning
of {\it LISA}'s mission.}
\label{fig:z1m1e5D}
\end{figure}

\begin{figure}
\epsfig{file = 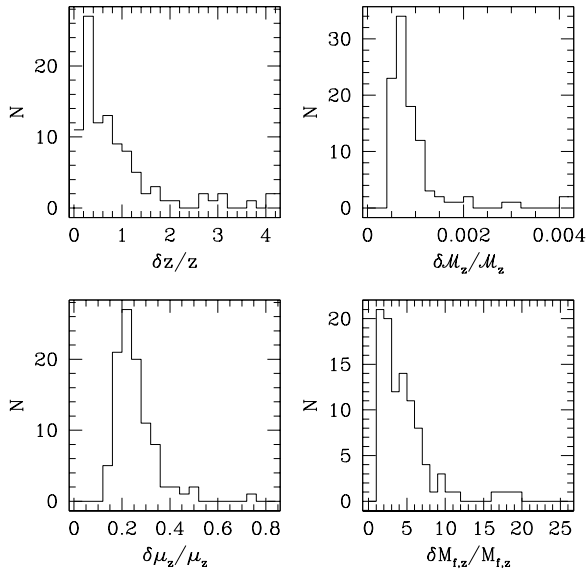, width = 8cm}
\caption{Distribution of errors in $z$, ${\cal M}_z$, $\mu_z$, and
$M_{f,z}$ for {\it LISA} measurement of a binary with $m_1 = m_2 =
10^4\,M_\odot$ at $z = 7$.  The typical inspiral SNR is about 45; the
ringdown SNR is around $0.7$.  Because of the weak ringdown waves, the
final mass is rather poorly determined.  In this case, measurement
error and cosmological parameter error contribute to the redshift
error about equally.  Setting the cosmological parameter errors to
zero in this case does not have a large effect.}
\label{fig:z7m1e4}
\end{figure}

The binaries chosen were placed at $z = 1,3,5,7,9$, and had masses
$m_1 = m_2 = 10^{3,4,5,6,7}\,M_\odot$.  For each redshift and mass, we
develop the distribution of parameter determination errors.  Two
examples of such distributions are illustrated in Figures
{\ref{fig:z1m1e5}} and {\ref{fig:z7m1e4}}.  Figure {\ref{fig:z1m1e5}}
shows the distribution of errors in $z$, ${\cal M}_z$, $\mu_z$, and
$M_{f,z}$ expected for measurements of a binary with $m_1 = m_2 =
10^5\,M_\odot$ at $z = 1$.  This is a case where most parameters are
measured rather well.  Note in particular the extreme precision with
which the redshifted chirp mass is determined: the distribution peaks
at $\delta{\cal M}_z/{\cal M}_z \sim 1.5\times 10^{-4}$.  The chirp
mass is measured so precisely because it most strongly determines the
inspiral rate, and hence has the greatest impact on the gravitational
wave phase evolution.  The reduced mass also impacts the inspiral, but
not as strongly, and hence is measured with less precision: the peak
in the distribution is at $\delta\mu_z/\mu_z\sim 2.5\times10^{-2}$.
Finally, note that the redshift errors peak at $\delta z/z\sim 15\%$.
This is the value expected when the redshift error is dominated by the
present uncertainty in our knowledge of cosmological parameters.  The
luminosity distance is actually determined far better than this,
indicating that improved knowledge of cosmological parameters will
greatly reduce errors in the redshift.  This is illustrated in Fig.\
{\ref{fig:z1m1e5D}}.  The top panel shows the distribution in
luminosity distance errors, and the bottom panel illustrates the
redshift error that would be achieved if we knew the cosmological
parameters exactly.  Comparing the two panels shows that $\delta
z/z\simeq \delta D/D$ when the cosmological parameters are known
accurately.  Both of these distributions peak near $1\%$ relative
error, and are largely confined to less than $10\%$ error.  The rule
of thumb ``$\delta z/z\simeq\delta D/D$ when $\delta(\mbox{cosmology})
= 0$'' holds more or less independent of redshift. As cosmological
parameters become better determined, our ability to determine the
redshift (and thus the masses) of coalescing binary black holes will
be greatly improved.

Table {\ref{tab:covar_z=1_m=1e5}} is a somewhat massaged
representation of the covariance matrix $\Sigma^{ab}$ for a typical
inspiral in this distribution --- diagonal components are actually the
mean error $\langle(\delta\theta^a)^2\rangle^{1/2}
=\sqrt{\Sigma^{aa}}$, off-diagonal components are the correlation
coefficient $c^{ab}$ defined in Eq.\ (\ref{eq:correlate}).  Note the
strong correlations between the luminosity distance and the position
and orientation angles.  These parameters are rather strongly
entangled since they set the amplitude of the measured waveform: the
angles through the detector response functions $F^+$ and $F^\times$
and the ratio of the wave's polarizations; the luminosity distance
through the overall amplitude ${\cal A}$ [cf.\ Eq.\
(\ref{eq:insp_amp})].  In order to make a good measurement of the
luminosity distance, we must determine the binary's sky position and
orientation very accurately.  The sky position and orientation are
encoded in the modulations induced by {\it LISA}'s orbital motion;
they are well determined when the waveform is subject to a large
amount of this motion-induced modulation.  Hence, luminosity distance
is only well determined when the waveform is significantly modulated
by the detector motion.  This accounts for the rather large tails
apparent in Figs.\ {\ref{fig:z1m1e5}} and {\ref{fig:z1m1e5D}}: though
most of the distribution for $\delta D/D$ (for example) is confined to
small error, some of the coalescences in the sample are measured with
much larger $\delta D/D$, up to $60\%$.  These large error
measurements occur when the inspiral is shorter than usual --- because
we randomly distribute the merger time during {\it LISA}'s mission,
some events occur very near the mission's start.  The position and
orientation angles are poorly determined for these short inspirals, so
the luminosity distance is also determined poorly.

\begin{figure}
\epsfig{file = 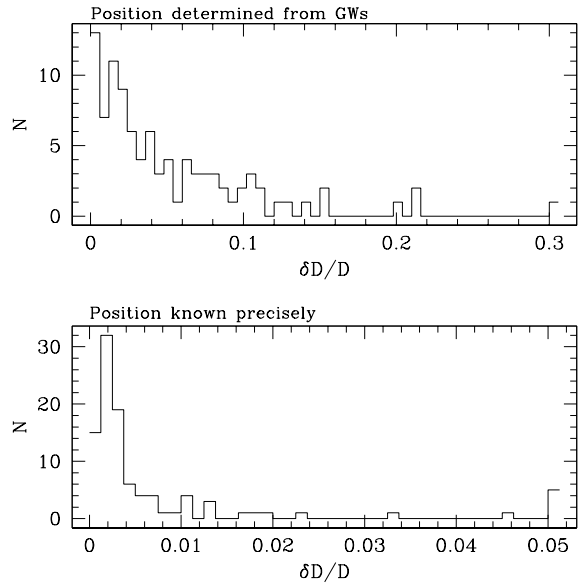, width = 8cm}
\caption{Comparison of measurement errors in luminosity distance
for {\it LISA} measurement of a binary with $m_1 = m_2 =
10^5\,M_\odot$ at $z = 1$.  The top panel shows the error distribution
when the sky position is determined using gravitational waves; the
bottom panel shows the distribution assuming the sky position is
measured with no error.  Because of strong correlations between the
luminosity distance and the sky position angles, improving the
accuracy with which the position is measured has a big impact on the
distance determination.  In this case, $D$ is measured with roughly an
order of magnitude less error.}
\label{fig:z1m1e5_compD}
\end{figure}

The impact of these correlations on the accuracy with which $D$ is
measured can be assessed by imagining that the binary's sky position
is measured with zero error.  This is a reasonable description of what
might be achieved by coordinated electromagnetic and
gravitational-wave measurements of binary coalescence --- for
instance, if the merger is accompanied by a gamma or x-ray flare.  The
electromagnetic measurement will likely determine the merger's sky
position much more accurately than can be done with gravitational
waves.  Figure {\ref{fig:z1m1e5_compD}} compares how well $D$ is
measured when the sky position is known precisely to the case of sky
position determined from the gravitational waves.  Knowledge of the
binary's sky position has an enormous impact, improving the accuracy
with which $D$ is measured by about an order of magnitude.

Note that Table {\ref{tab:covar_z=1_m=1e5}} also shows strong
correlations among the mass and spin parameters, ${\cal M}_c$, $\mu$,
$\beta$, and $\sigma$.  This is a well-known feature of measurement
with post-Newtonian templates, arising because the impact of the
various parameters upon the phase evolution are not strongly different
from one another.  For further discussion, see Cutler \& Flanagan
(1994) and Poisson \& Will (1996).

Figure {\ref{fig:z7m1e4}} shows the distribution of measurement errors
when $m_1 = m_2 = 10^4\,M_\odot$ and $z = 7$.  In all cases, the
distributions peak at larger error values than when $m_1 = m_2 =
10^5\,M_\odot$ and $z = 1$.  This isn't too surprising since these
sources are much fainter and thus harder to measure.  The redshift
determination in particular is quite a bit worse, so that the
mass-redshift degeneracy will be broken rather less accurately for
these holes.  Table {\ref{tab:covar_z=7_m=1e4}} shows the parameter
errors and correlations for a typical coalescence in this set.
Because of the signal's weakness, the measurement errors (diagonal
components of the matrix) are rather larger than in the case $z = 1$,
$m_1 = m_2 = 10^5\,M_\odot$.  However, the correlations (off-diagonal
components) are not much different.  This is typical: correlations
between inspiral parameters do not depend strongly on signal-to-noise
ratio, though the inspiral time can have a big effect.

Rather than show distribution histograms for all the remaining cases
that we examined, we summarize their contents in Tables
{\ref{tab:z=1}} -- {\ref{tab:z=9}}.  These tables give the ``most
likely'' error values --- the errors found at the peaks of the
measurement distribution.  The reader should bear in mind that the
distributions from which these values were taken also have long,
high-error tails, as in the Figures.

Several interesting features can be seen across the tables.  Only
Table {\ref{tab:z=1}} includes data for the merger of holes with $m_1
= m_2 = 10^7\,M_\odot$.  At redshifts higher than $z = 1$, the
inspiral signal from such binaries is radiated at frequencies entirely
below $10^{-4}\,{\rm Hz}$, out of {\it LISA}'s band.  Indeed, even at
$z = 1$ this binary barely radiates in band, lasting less than 2 hours
before merger.  The signal is so short that little inspiral phase
accumulates, so $D$, ${\cal M}_z$, and $\mu_z$ are determined very
poorly.  The ringdown, by contrast, is quite strong, so $M_{f,z}$ is
measured with very good accuracy.

The tables show that, in general, measurement accuracy degrades with
increasing redshift.  This isn't surprising since such sources are
fainter.  However, at each redshift, there are some sources and
parameters that can be measured with at least moderate precision.  At
low redshift, there is a fairly broad range of masses in which at
least two of the masses (${\cal M}_z$, $\mu_z$, $M_{f,z}$) can be
determined with good precision, and in some cases all three are well
measured.  Out to $z = 5$, there are cases for which $\delta D/D
<\delta z/z$, indicating that in those cases the redshift distribution
is skewed high because of the present error in $h_0$ and
$\Omega_\Lambda$.  Setting the cosmological parameter error to zero,
we find $\delta z/z\simeq\delta D/D$, as expected.  Improved knowledge
of cosmological parameters will have a big impact on measurement in
those cases.  At redshifts $z > 5$, $\delta z$ becomes dominated by
error in $D$ --- improved knowledge of $h_0$ and $\Omega_\Lambda$ will
not have much effect.

Finally, note that the tables confirm the trend discussed above that
$D$ (and hence $z$) is not well determined for very short measurement
times: if the inspiral does not last ``long enough'', the
motion-induced modulation of the waveform is not sufficient to
determine the source's sky position and orientation very accurately.
As a consequence, $D$ is poorly determined.  A rough necessary
condition to measure $D$ well seems to be that the {\it LISA}
constellation must move through a radian or so of its orbit.  This is
{\it not}, however, a sufficient condition --- $D$ can be determined
poorly from long inspirals if the signal is too weak.

In all cases except for the largest masses, the best determined
parameter is the redshifted chirp mass.  It is often measured with
precision $\delta{\cal M}_z/{\cal M}_z\lesssim 0.1\%$ or better, even
from sources with $z\sim9$.  {\it LISA}'s ability to break the
mass-redshift degeneracy for the chirp mass will therefore be limited
by redshift error: we will measure the {\it actual} chirp mass of
distant coalescences to within $15-30\%$ assuming present cosmological
parameters, and perhaps as well as $5-30\%$ when {\it LISA} actually
flies.  The reduced mass is often measured with a precision of about
$10\%$ or better, and so provides useful additional information.  The
final mass of the system is only determined from more massive systems,
since the ringdown waves are relatively high frequency.  In those
cases, it can be measured with accuracy $\delta
M_{f,z}/M_{f,z}\lesssim 5\%$.

\section{Summary and conclusion}
\label{sec:conclude}

By combining gravitational-wave measurements with information about
cosmological parameters, {\it LISA} will be able to measure the
redshift of coalescing binaries with moderate precision (relative
error of $15-30\%$ using present uncertainties in cosmological
parameters, perhaps $5-30\%$ by the time that {\it LISA} flies).  The
redshifted chirp mass is typically measured far more accurately than
this, and in many cases, either the binary's reduced mass or the final
mass of the remnant black hole produced when the binary merges can be
measured with a precision of $5-20\%$.  This precision will allow {\it
LISA} to untangle mass and redshift, making it possible to track the
merger history of massive black holes in the universe.

These measurements work best for signals whose frequency range lies in
{\it LISA}'s band of maximum sensitivity.  From Tables {\ref{tab:z=1}}
-- {\ref{tab:z=9}}, we see that the best measurement sensitivities are
for systems that have $M_z \sim 10^5\,M_\odot$ or so.  This isn't too
surprising: recall that inspiral ends when the bodies are separated by
a distance $r \sim 6 M$, and that the gravitational-wave frequency at
that point is
\begin{equation}
f_{\rm GW}\simeq 0.04\,{\rm Hz}\left({10^5\,M_\odot\over M_z}\right)
\end{equation}
[cf.\ Eq.\ (\ref{eq:f_end_insp})].  This frequency is right about
where {\it LISA}'s sensitivity begins to degrade due to high frequency
noise.  When $M_z\sim10^5\,M_\odot$ most of the inspiral signal
accumulates in a band where {\it LISA} has very good sensitivity.  A
good rule of thumb seems to be that {\it LISA} will measure at least
two of the redshifted mass combinations ${\cal M}_z$, $\mu_z$, and
$M_{f,z}$ with good precision when the total redshifted system mass is
within a factor of ten to twenty of $10^5\,M_\odot$.  Even outside
that range, at least one mass (either the chirp mass or the final
mass) can be measured well.

To measure the redshift precisely, we first must determine the
luminosity distance.  This correlates most strongly with the amount of
time over which the inspiral signal is measured.  Very short inspirals
do not experience enough detector-motion-induced modulation of the
gravitational waveform to pin down a source's location on the sky very
accurately, and as a consequence the luminosity distance can be poorly
determined.  This is the reason the largest systems in our sample have
such poor precision in $D$ and $z$: they enter {\it LISA}'s band
already very close to merging, and quickly evolve to merger.  The
distance is determined well when {\it LISA} moves through at least a
radian or so of its orbit, corresponding to at least 2 months of
observation.

This analysis is essentially just a first cut, proof-of-principle
demonstration of how cosmological information and gravitational-wave
measurements can be combined to study the merger history of massive
black holes.  We have made several simplifying assumptions that, if
lifted, may modify some of our conclusions.  One example is our use of
the restricted post-Newtonian approximation.  In analyzing the
inspiral, we have thrown out all but the strongest, quadrupole
harmonic of the binary's orbit.  Other harmonics contribute as well,
though, and should be measurable by {\it LISA}.  Because the amplitude
of each harmonic depends on source parameters (particularly the
binary's inclination angle) in a different way, measuring multiple
harmonics could greatly improve the distance determination by breaking
degeneracies between $D$ and the source orientation
angles\footnote{The author is very grateful to Ron Hellings for
bringing this point to his attention.}  (apparent in the large
correlation coefficients shown in Tables {\ref{tab:covar_z=1_m=1e5}}
and {\ref{tab:covar_z=7_m=1e4}}).

A better understanding of the merger epoch could also improve
parameter estimation.  The merger waves are, by definition, all of the
radiation emitted at frequencies between the end of inspiral and the
ringdown.  In some cases, these waves are likely to be very strong and
interesting, particularly if the black holes in the binary are rapidly
rotating {\cite{bbhI,pricewhelan}}.  They will probably be rather
short duration (perhaps 10-20 cycles at $f \sim 0.05\,{\rm Hz}$ when
$M_z \sim 10^5\,M_\odot$), and so provide essentially no information
about the source location and distance.  But they could give some
information about the binary's masses and spins, which could break
some degeneracies and improve the accuracy of parameters measured
during the inspiral.

We have used a rather simple, smooth power-law fit to {\it LISA}'s
projected instrumental noise curve.  This is adequate for first-cut
estimates, but is in principle wrong, particularly at high frequencies
when the radiation wavelength is shorter than {\it LISA}'s arms.
Following the analysis of Larson, Hiscock, \& Hellings (2000) it is
straightforward to describe {\it LISA}'s response to a
gravitational-wave more precisely.  In practice, this will require a
moderate amount of effort, particularly since {\it LISA}'s response
function and noise curve turn out to depend upon a source's sky
position.  This response will put additional position-dependent
structure into a particular measurement, which could be used to
further improve positional accuracy.  This merits some detailed
investigation.

A source of systematic error that has been neglected in this analysis
is gravitational lensing.  Gravitational waves are lensed exactly as
electromagnetic radiation is lensed, and this will impact the accuracy
with which parameters are measured.  Since {\it LISA} will be watching
binary coalescence events out to large redshift, we expect that weak
lensing will be common; if the number of mergers is large, there may
also be some strongly lensed mergers.  Lensing will have little or no
effect on the determination of the redshifted masses, since they are
measured from the signal's phase evolution, but will impact the wave's
amplitude.  A lens with large skew could also change the relative
magnitude of the $+$ and $\times$ polarizations.  The source's
orientation and sky position could thus be determined incorrectly,
which may skew the measurement of $D$, and thence the determination of
$z$ {\cite{markovic}}.  

Finally, it would be very useful to tie these estimates more closely
to hierarchical structure formation scenarios (see, e.g., Menou,
Haiman, \& Narayanan 2001), and to understand how much {\it LISA}
could impact such models.  It would also be valuable to understand the
rate of mergers at large redshift [in analogy with the analysis of
Menou, Haiman, \& Narayanan (2001), which estimates rates out to $z
\sim 5$].  If this rate turns out to be {\it too} high, binary black
hole mergers might actually be a confusion limited source of noise
{\cite{phinney_mnras1}} rather than an opportunity to study structure
formation.

I thank Ron Hellings, Daniel Holz, Shane Larson, and Sterl Phinney for
many useful discussions.  I am also grateful to the astrophysics and
relativity group of Cornell University and the Center for
Gravitational Physics and Geometry at Penn State, where a portion of
the code used in this analysis was written.  This research was
supported by NSF Grant PHY--9907949.

{}

\begin{table*}
\begin{minipage}{16cm}
\caption{Measurement accuracy and correlations for coalescences at
$z = 1$ with $m_1 = m_2 = 10^5\,M_\odot$.  Diagonal elements in this
matrix are the mean error expected in the parameter; off-diagonal
elements are the correlation coefficient $c^{ab}$.  Elements
containing $\cdot$ can be found by symmetry.  The errors in $\phi_L$,
$\phi_S$, and $\phi_c$ are in radians; those in $t_c$ are in units of
$10^4$ seconds.  All other entries are dimensionless.}
\begin{tabular}{l|ccccccccccc}
 & $\ln D$ & $\mu_L$ & $\mu_S$ & $\phi_L$ & $\phi_S$ & $t_c$ &
$\phi_c$ & $\ln{\cal M}_z$ & $\ln\mu_z$ & $\beta$ & $\sigma$\\
\hline
$\ln D$ & 0.0095  & 0.8969  & 0.8258  & 0.6692  & 0.8185  & -0.0532 & 0.3224  & 0.1608  & -0.2196 & 0.2182  & -0.2076\\
$\mu_L$  & $\cdot$ & 0.0114  & 0.8215  & 0.2972  & 0.8881  & -0.1140 & 0.2273  & -0.0143 & -0.0625 & 0.0502  & -0.1048\\
$\mu_S$  & $\cdot$ & $\cdot$ & 0.0006  & 0.4474  & 0.7591  & -0.1786 & 0.2526  & 0.0823  & -0.1402 & 0.1353  & -0.1483\\
$\phi_L$ & $\cdot$ & $\cdot$ & $\cdot$ & 0.0362  & 0.3280  & 0.0777  & 0.3351  & 0.3884  & -0.3883 & 0.4063  & -0.2879\\
$\phi_S$ & $\cdot$ & $\cdot$ & $\cdot$ & $\cdot$ & 0.0051  & -0.0802 & 0.2074  & 0.0067  & -0.0703 & 0.0612  & -0.0100\\
$t_c$    & $\cdot$ & $\cdot$ & $\cdot$ & $\cdot$ & $\cdot$ & 0.0002  & 0.8837  & 0.7522  & -0.8287 & 0.7980  & -0.9189\\
$\phi_c$ & $\cdot$ & $\cdot$ & $\cdot$ & $\cdot$ & $\cdot$ & $\cdot$ & 0.7179  & 0.8691  & -0.9473 & 0.9247  & -0.9918\\
$\ln{\cal M}_z$ &$\cdot$ & $\cdot$ & $\cdot$ & $\cdot$ & $\cdot$ & $\cdot$ & $\cdot$ & 0.0001  & -0.9798 & 0.9904  & -0.8828\\
$\ln\mu_z$ & $\cdot$ & $\cdot$ & $\cdot$ & $\cdot$ & $\cdot$ & $\cdot$ & $\cdot$ & $\cdot$ & 0.0334  & -0.9976 & 0.9556\\
$\beta$  & $\cdot$ & $\cdot$ & $\cdot$ & $\cdot$ & $\cdot$ & $\cdot$ & $\cdot$ & $\cdot$ & $\cdot$ & 0.4171  & -0.9331\\
$\sigma$ & $\cdot$ & $\cdot$ & $\cdot$ & $\cdot$ & $\cdot$ & $\cdot$ & $\cdot$ & $\cdot$ & $\cdot$ & $\cdot$ & 0.3251 \\
\hline
\end{tabular}
\label{tab:covar_z=1_m=1e5}
\end{minipage}
\end{table*}

\begin{table*}
\begin{minipage}{16cm}
\caption{Measurement accuracy and correlations for coalescences at
$z = 7$ with $m_1 = m_2 = 10^4\,M_\odot$.}
\begin{tabular}{lccccccccccc}
 & $\ln D$ & $\mu_L$ & $\mu_S$ & $\phi_L$ & $\phi_S$ & $t_c$ &
$\phi_c$ & $\ln{\cal M}_c$ & $\ln\mu_z$ & $\beta$ & $\sigma$\\
\hline
$\ln D$  & 0.2036  & -0.2030 & 0.8774  & -0.6732 & -0.6861 & -0.3873 & -0.2672 & -0.4736 & 0.3933  & -0.4331 & 0.2200\\
$\mu_L$  & $\cdot$ & 0.6189  & -0.5836 & -0.4715 & -0.4980 & -0.7733 & 0.0455  & 0.0608  & -0.0245 & 0.0258  & -0.0177\\
$\mu_S$  & $\cdot$ & $\cdot$ & 0.3647  & -0.4104 & -0.3818 & -0.0090 & -0.2387 & -0.4250 & 0.3413  & -0.3758 & 0.1907\\
$\phi_L$ & $\cdot$ & $\cdot$ & $\cdot$ & 0.7262  & 0.9927  & 0.9066  & 0.2173  & 0.3933  & -0.3429 & 0.3792  & -0.1863\\
$\phi_S$ & $\cdot$ & $\cdot$ & $\cdot$ & $\cdot$ & 0.4652  & 0.9261  & 0.2182  & 0.3898  & -0.3412 & 0.3771  & -0.1861\\
$t_c$    & $\cdot$ & $\cdot$ & $\cdot$ & $\cdot$ & $\cdot$ & 0.0083  & 0.1932  & 0.2900  & -0.2768 & 0.2992  & -0.1778\\
$\phi_c$ & $\cdot$ & $\cdot$ & $\cdot$ & $\cdot$ & $\cdot$ & $\cdot$ & 4.0390  & 0.9008  & -0.9562 & 0.9354  & -0.9960\\
$\ln{\cal M}_z$ &$\cdot$ & $\cdot$ & $\cdot$ & $\cdot$ & $\cdot$ & $\cdot$ & $\cdot$ & 0.0005  & -0.9848 & 0.9931  & -0.9012\\
$\ln\mu_z$ & $\cdot$ & $\cdot$ & $\cdot$ & $\cdot$ & $\cdot$ & $\cdot$ & $\cdot$ & $\cdot$ & 0.1479  & -0.9977 & 0.9589\\
$\beta$  & $\cdot$ & $\cdot$ & $\cdot$ & $\cdot$ & $\cdot$ & $\cdot$ & $\cdot$ & $\cdot$ & $\cdot$ & 2.1417  & -0.9373\\
$\sigma$ & $\cdot$ & $\cdot$ & $\cdot$ & $\cdot$ & $\cdot$ & $\cdot$ & $\cdot$ & $\cdot$ & $\cdot$ & $\cdot$ & 1.9260 \\
\hline
\end{tabular}
\label{tab:covar_z=7_m=1e4}
\end{minipage}
\end{table*}

\begin{table*}
\begin{minipage}{16cm}
\caption{Summary of measurement accuracies for binary black
hole coalescences at $z = 1$.  The redshift error $\delta z_1$ assumes
$\delta h_0 = 0.1$, $\delta\Omega_\Lambda = 0.1$; the error $\delta
z_2$ assumes the cosmological parameters are known perfectly.}
\begin{tabular}{lccccccccc}
\hline
{$m_1 ( = m_2)$} &
{$\rho_{\rm insp}$} &
{$T_{\rm insp}$} &
{$\rho_{\rm ring}$} &
{$\delta D/D$} &
{$\delta z_1/z$} &
{$\delta z_2/z$} &
{$\delta{\cal M}_z/{\cal M}_z$} &
{$\delta\mu_z/\mu_z$} &
{$\delta M_{f,z}/M_{f,z}$}\\
\hline
$10^3\,M_\odot$ &
20 & 575 days & $10^{-3}$ & 0.08 & 0.15 & 0.05 & $5\times10^{-5}$ & 0.05 & 2500 \\
$10^4\,M_\odot$ &
150 & 550 days & 0.25 & 0.05 & 0.15 & 0.04 & $5\times10^{-5}$ & 0.03 & 10 \\
$10^5\,M_\odot$ &
1000 & 430 days & 60 & 0.02 & 0.15 & 0.02 & $1\times10^{-4}$ & 0.04 & 0.03 \\
$10^6\,M_\odot$ &
200 & 15 days & 3500 & 0.2 & 0.2 & 0.2 & $5\times10^{-3}$ & 0.5 & $5\times10^{-4}$ \\
$10^7\,M_\odot$ &
40 & 100 minutes & 612 & 70 & 70 & 70 & 150 & 300 & $4\times10^{-3}$ \\
\hline
\end{tabular}
\label{tab:z=1}
\end{minipage}
\end{table*}

\begin{table*}
\begin{minipage}{16cm}
\caption{Summary of measurement accuracies for binary black
hole coalescences at $z = 3$.}
\begin{tabular}{lccccccccc}
\hline
{$m_1 ( = m_2)$} &
{$\rho_{\rm insp}$} &
{$T_{\rm insp}$} &
{$\rho_{\rm ring}$} &
{$\delta D/D$} &
{$\delta z_1/z$} &
{$\delta z_2/z$} &
{$\delta{\cal M}_z/{\cal M}_z$} &
{$\delta\mu_z/\mu_z$} &
{$\delta M_{f,z}/M_{f,z}$}\\
\hline
$10^3\,M_\odot$ &
10 & 575 days & $10^{-3}$ & 0.2 & 0.2 & 0.2 & $1\times10^{-4}$ & 0.08 & 1000 \\
$10^4\,M_\odot$ &
75 & 530 days & 0.35 & 0.1 & 0.15 & 0.1 & $5\times10^{-4}$ & 0.1 & 5 \\
$10^5\,M_\odot$ &
400 & 200 days & 85 & 0.1 & 0.15 & 0.1 & $1\times10^{-3}$ & 0.3 & 0.02 \\
$10^6\,M_\odot$ &
70 & 5 days & 1300 & 0.8 & 0.8 & 0.8 & $1.5\times10^{-2}$ & 0.6 & $1\times10^{-3}$ \\
\hline
\end{tabular}
\label{tab:z=3}
\end{minipage}
\end{table*}

\begin{table*}
\begin{minipage}{16cm}
\caption{Summary of measurement accuracies for binary black
hole coalescences at $z = 5$.}
\begin{tabular}{lccccccccc}
\hline
{$m_1 ( = m_2)$} &
{$\rho_{\rm insp}$} &
{$T_{\rm insp}$} &
{$\rho_{\rm ring}$} &
{$\delta D/D$} &
{$\delta z_1/z$} &
{$\delta z_2/z$} &
{$\delta{\cal M}_z/{\cal M}_z$} &
{$\delta\mu_z/\mu_z$} &
{$\delta M_{f,z}/M_{f,z}$}\\
\hline
$10^3\,M_\odot$ &
9 & 540 days & $2\times10^{-3}$ & 0.25 & 0.3 & 0.25 & $2\times10^{-4}$ & 0.1 & 1000 \\
$10^4\,M_\odot$ &
55 & 560 days & 0.6 & 0.2 & 0.2 & 0.2 & $2\times10^{-4}$ & 0.2 & 4 \\
$10^5\,M_\odot$ &
250 & 100 days & 100 & 0.15 & 0.2 & 0.15 & $3\times10^{-3}$ & 0.5 & 0.02 \\
$10^6\,M_\odot$ &
30 & 2.5 days & 350 & 2 & 2 & 2 & 0.05 & 0.6 & 0.01 \\
\hline
\end{tabular}
\label{tab:z=5}
\end{minipage}
\end{table*}

\begin{table*}
\begin{minipage}{16cm}
\caption{Summary of measurement accuracies for binary black
hole coalescences at $z = 7$.}
\begin{tabular}{lccccccccc}
\hline
{$m_1 ( = m_2)$} &
{$\rho_{\rm insp}$} &
{$T_{\rm insp}$} &
{$\rho_{\rm ring}$} &
{$\delta D/D$} &
{$\delta z_1/z$} &
{$\delta z_2/z$} &
{$\delta{\cal M}_z/{\cal M}_z$} &
{$\delta\mu_z/\mu_z$} &
{$\delta M_{f,z}/M_{f,z}$}\\
\hline
$10^3\,M_\odot$ &
9 & 540 days & $3\times10^{-3}$ & 0.3 & 0.3 & 0.3 & $2.5\times10^{-4}$ & 0.15
& 500 \\
$10^4\,M_\odot$ &
46 & 560 days & 0.7 & 0.25 & 0.25 & 0.25 & $7\times10^{-4}$ & 0.25 & 1.5 \\
$10^5\,M_\odot$ &
150 & 65 days & 120 & 0.5 & 0.5 & 0.5 & $5\times10^{-3}$ & 0.6 & 0.015 \\
$10^6\,M_\odot$ &
24 & 1.5 days & 100 & 6 & 6 & 6 & 0.1 & 0.7 & 0.03 \\
\hline
\end{tabular}
\label{tab:z=7}
\end{minipage}
\end{table*}

\begin{table*}
\begin{minipage}{16cm}
\caption{Summary of measurement accuracies for binary black
hole coalescences at $z = 9$.}
\begin{tabular}{lccccccccc}
\hline
{$m_1 ( = m_2)$} &
{$\rho_{\rm insp}$} &
{$T_{\rm insp}$} &
{$\rho_{\rm ring}$} &
{$\delta D/D$} &
{$\delta z_1/z$} &
{$\delta z_2/z$} &
{$\delta{\cal M}_z/{\cal M}_z$} &
{$\delta\mu_z/\mu_z$} &
{$\delta M_{f,z}/M_{f,z}$}\\
\hline
$10^3\,M_\odot$ &
8 & 580 days & $4\times10^{-3}$ & 0.5 & 0.5 & 0.5 & $3\times10^{-4}$ & 0.16
& 500 \\
$10^4\,M_\odot$ &
43 & 530 days & 0.9 & 0.3 & 0.3 & 0.3 & $7\times10^{-4}$ & 0.25 & 1.5 \\
$10^5\,M_\odot$ &
100 & 45 days & 130 & 1 & 1 & 1 & $7\times10^{-3}$ & 0.6 & 0.015 \\
$10^6\,M_\odot$ &
17 & 1 day & 40 & 20 & 20 & 20 & 0.6 & 1.5 & 0.05 \\
\hline
\end{tabular}
\label{tab:z=9}
\end{minipage}
\end{table*}

\label{lastpage}
\bsp

\end{document}